# Phonon Activity and a Possible Lattice Arrangement in the Intermediate Phase of YVO$_3$

Néstor E. Massa[*]
Laboratorio Nacional de Investigación y Servicios en Espectroscopía Optica, Centro CEQUINOR-Departamento de Química and Departamento de Física, Universidad Nacional de La Plata, C.C. 962, 1900 La Plata, Argentina,

Cínthia Piamonteze
Laboratório Nacional de Luz Síncrotron, CP 6192, 13083-970 and Instituto de Física, Universidade Estadual de Campinas, 13083-970, Campinas, São Paulo, Brazil,

Hélio C. N. Tolentino
Laboratório Nacional de Luz Síncrotron, CP 6192, 13083-970, Campinas, São Paulo, Brazil,

and

José Antonio Alonso, María Jesús Martínez-Lope, María Teresa Casais
Instituto de Ciencia de Materiales de Madrid, Consejo Superior de Investigaciones Científicas, Cantoblanco, E-28049 Madrid, Spain.

---

- e-mail :nem@dalton.quimica.unlp.edu.ar

# ABSTRACT

We show that in YVO$_3$ new hard phonons gradually become zone center active below ~210 K verifying that a lattice phase transition takes place at about that temperature. The Debye-Waller factor found from EXAFS measurements show an additional disorder below150K seen as a departure from the thermal behavior. We bring up the similarities of  the intermediate phase to known results belonging to incommensurate systems that here we attribute to faulty orbital stacking with possible domain formation set below 116K, the onset temperature of the G-type antiferromagnetism. We found support for this interpretation in the inverted lambda shapes in the measured heat capacity at the phase transitions and in the overall temperature dependence of the reported new hard phonons as compared with those for known incommensurate dielectrics.

A plethora of oxides with perovskite structure have been described in the last decades, with fascinating electrical or magnetic properties ranging from high Tc superconductivity to colossal magnetoresistence or ferroelectricity. Among them, $RVO_3$ vanadates (R= rare earth and Y) are canted spin antiferromagnets[1] conforming a family in which anharmonicites, lattice instabilities, Jahn-Teller distortions, intermixed with complex magnetic arrangements, are expected to play a no negligible role in the phonon response. It is known that transition metal oxides, in general, are materials predisposed to easily contain imperfections in a nominally pure phase. Then, understanding orbital degrees of freedom and short range anisotropies in confined lattice deformations will clarify the role triggering instabilities yielding to distortions that might be responsible for a lower symmetry phase transition.

Here we focus our attention on $YVO_3$, a compound regarded as a Mott-Hubbard insulator.[2] It is an orthorhombic distorted perovskite belonging at room temperature to the *Pbnm* space group where the rare earth size favors the $GdFeO_3$ structure, a distortion that may be also be incremented by contributing Jahn-Teller effects. The $V^{3+}$ ions, at octahedral sites, show a $3d^2$ electronic configuration with spin S=1, where the $t_{2g}$ orbitals, orthogonal in the cubic phase, are filled with two electrons.

The magnetic properties of YVO$_3$ have been exhaustively studied.[3] Multiple and reversible sign changes in the magnetization also result in a temperature interval in which the magnetic moment is oriented in a direction opposite to the applied magnetic field. The passage from a paramagnetic regime to a nearly antiferromagnetic (AFM) below $T_{N1}$ ~116 K is achieved by canting yielding a C-type AFM structure that holds down to ~77 K.[4] Remarkably, the net magnetization in single crystals increases to a maximum and then decrease to reach negative values at about 95 K. It is claimed that spin reorientations at these temperatures are not accompanied by structural changes.[3] At temperatures close to 77 K,~$T_{N2}$, where the structural transition into a monoclinic distortion takes place, the net moment switches again to positive values into a C-type AFM that only consolidates at ~50 K.[5] It is found that the distinct magnetic orderings undergo simultaneous changes of orbital ordering and show hysteric behavior. G-type orbital ordering has also been reported below 77 K by resonant x-ray scattering at the vanadium K edge.[6]

Then, since it is known that in oxides the lattice plays a determinant role in intrinsic properties, it is pertinent a phonon study as a function of temperature in order to relate the findings to local structure deviations. This might correlate with the magnetic anomalies commented in the preceding paragraphs by relating short range features to temperature dependent defective orbital mixing and octahedral distortion implying separate formation of nanoclusters in the intermediate phase of YVO$_3$. Our studies are also partially motivated by the recently reported failure of first principle calculations in describing the intermediate phase of YVO$_3$ as it has been accounted in literature.[7]

Here, we report infrared temperature dependent measurements of polycrystalline samples of YVO$_3$ (ref. 8) and further interpret those spectra with EXAFS (Extended X-ray Absorption Fine Structure) studies of V-K edge.

Our temperature dependent spectra were run using a FT-IR Bruker 113v interferometer in the 30-10000 cm$^{-1}$ frequency range with samples mounted on a cold finger of an OXFORD DN 1754 cryostat. The temperature stability was 0.1 K. Reflectivity measurements were made from pure polycrystalline pellets, while semitransparent CsI and polyethylene 1-cm-diameter pellets, embedded randomly with YVO$_3$ microcrystals, were also made in order to determine the transmission spectra. They have 2 and 1 cm$^{-1}$ resolution respectively. For reflectivity measurements, a gold mirror was used as a 100% reference.

EXAFS measurements were recorded using the XAS beam line installed at the 1.37 GeV storage ring of the Laboratório Nacional de Luz Síncroton at Campinas, Brazil.[9] For this purpose, portions of the pure samples used in our infrared reflectivity were reduced to fine powder in an agate mortar. The fine-ground powder, washed with alcohol, was then spread onto nitro-cellulose filter, resulting after evaporation, in a uniformly thick film.

Single phase $YVO_3$ samples were prepared in polycrystalline form by soft-chemistry procedures. $Y_2O_3$ and $NH_4VO_3$ were dissolved in citric acid the citrate solutions were dried and decomposed in air at 700 ºC leading to homogeneous and reactive precursor powers. The precursors were subsequently annealed under reducing flows ($H_2/N_2$) at temperatures of 1100-1160 ºC. We have also prepared $YVO_3$ by solid state reactions under 20 kbar and 1000 ºC. Both set of samples yielded the same results.

Primary characterization of our samples has been done by X-ray diffraction at room temperature. In addition, neutron powder diffraction experiments were performed at the D2B high-resolution diffractometer at the Institut Laue Langevin, Grenoble. The patterns refined by the Rietveld method yielded at room temperature the conventional P*bnm*-$D_{2h}^{16}$ space group while at 2 K a best fit was achieved in the monoclinic P21/n space group with two different octahedral environments. It is important, however, to note that although in overall our structural results agree with those by Blake et al[10] we also share their concerns on diffraction data acquisition and analysis particularly in the 200 K to 77 K temperature range.

The room temperature infrared reflectivity spectra of YVO$_3$ are identifiable with those of any other GdFeO$_3$ distorted perovskite.[11,12] At room temperature there are nine well defined bands that correspond to lattice, antisymmetric and symmetric stretching modes of the perovskite lattice centered at ≈200 cm$^{-1}$, ≈400 cm$^{-1}$ and ≈650 cm$^{-1}$ respectively (inset, figure 1). They may be considered as envelops of all room temperature infrared allowed modes and are easily followed down to 77 K. In addition, new weak phonon modes with onset at ≈210 K signal, as a substructure, a lattice phase transition. The most prominent new hard phonons gradually emerging from a typical lattice band shoulder are shown in figure 1. The lattice instability, although weak is gradually enhanced as the temperature is lowered toward 77 K through $T_{N1}$≈116 K, the temperature of the reported C-type antiferromagnetic ordering, and it is associated with weak reflections violating the P*bnm* symmetry reported by Blake et al on cooling at about 200 K.[10]

We estimated phonon frequencies, table 1, using a standard multioscillator dielectric simulation fit on our reflectivity spectra.[11] We analyzed the reflectivity spectra for each temperature simulating infrared-active features with damped Lorentzian oscillators in a classical formulation of the dielectric function. The expression of the dielectric function is given by

$$\varepsilon(\omega) = \varepsilon_\infty \prod_j \left[ \frac{\omega_{jLO}^2 - \omega^2 + i\gamma_{jLO}\omega}{\omega_{jTO}^2 - \omega^2 + i\gamma_{jTO}\omega} \right], \quad (1)$$

where $\varepsilon_\infty$ is the high-frequency dielectric function; $\omega_{jLO}$ and $\omega_{jTO}$ are the longitudinal and transverse jth optical frequencies with damping constants $\gamma_{jLO}$ and $\gamma_{jTO}$ respectively.

We also calculated the jth oscillator strength $S_j$ as

$$S_j = \omega_{jTO}^{-2} \frac{\prod_k \left(\omega_{kLO}^2 - \omega_{jTO}^2\right)}{\prod_{k \neq j} \left(\omega_{kTO}^2 - \omega_{jTO}^2\right)}. \quad (2)$$

The new zone center optical activity means a net Brillouin zone folding between 210 and 77 K. Consequently, in agreement with conclusions by Ren et al[3], we do not observe new infrared active modes as the lattice lower symmetry is already present above $T_{N1}$, changing continuously through it. The dielectric simulation analysis yields phonon strengths (those in bold letters in table 1 are plotted in figure 2) that allows identifying by optical means the temperature at which that phase transition takes place. Phonon intensity bands below ~210 K seem to obey a very weak power law.

The main phonon lattice band at ~160 cm$^{-1}$ may be visualized originating from Y ions moving against the VO$_6$ octahedra and then its behavior is likely related to its anomalous coordinate temperature dependence reported while the orbital ordering takes place as the sample cools down through ~210 K. This is the temperature for the Jahn-Teller distortion onset with two arrangements of alternating long and short V-O bonds forming two layers.[9]

In addition, that phonon activity is also mirrored by some internal mode band changes at higher frequencies that although are not as well infrared resolved, reflects on octahedral distortion in the intermediate phase.

The vibrational and structural findings are also supported by temperature dependent EXAFS results. Figure 3 shows the Fourier Transform amplitude (FT) of the EXAFS signal measured around V-K edge, which represents the pseudo-RDF (radial distribution function) around V atoms. It is important to remark that the R-axis does not corresponds to the atomic separation because a phase shift correction, intrinsic to the EXAFS analysis, was not taken into account yet. If there is no structural modification, the expected behavior for the FT curves is decreasing amplitude as temperature increases due to the thermal damping. This monotonous decrease is seen, for instance, for the metal-to-metal bonding (V-V). Considering the first peak of the FT curves, which correspond to the V-O neighbor shell, it is seen an anomalous behavior for the data measured below 150K. In order to quantify the amplitude variation in function of temperature, the V-O peak was analyzed considering three fixed V-O bond distances (inset, figure 4) using as a model depart the parameters previously found by Blake et al.[10] from neutron diffraction measurements. The Debye-Waller factor was maintained the same for the three distances and, therefore, it represents the disorder averaged over all directions relative to the model. The relative Debye-Waller factor ($\Delta\sigma^2$) obtained from this analysis[13] plotted against temperature (figure 4) shows a monotonic decrease with temperature down to 100K. Below 150K it is clearly observed an increase in $\sigma^2$ with a maximum at about 50 K, the onset of AFM ordering in the low temperature monoclinic phase.[5] Since the Debye-Waller factor measured by EXAFS takes into account both thermal and structural disorder, such increase, figure 4, is ascribed to the appearance of V local structural disorder. =

In a perovskite lattice with transition metal ions there are two Jahn-Teller (JT) competing distortions, a- and -d-type with the elongation/shortening within the basal plane. They differ only in the way they are stacked.[14] We may understand the above scenario and the gradual increment in phonon intensity as due to a progressively more pronounced $GdFeO_3$ distortion driven by the octahedra tilting and rotation. As Mizokawa et al[15] point out that at ~210 K, cooling toward 116 K, the system starts with energies of a C-type AFM, compatible with an a-type JT distortion, close to be degenerated to one for a G-type AFM-*d*-type JT distortion and thus, having net lattice distortion nearly suppressed at these relative higher temperatures. Below 116 K, a C-type antiferromagnetism favors an *a*-type JT distortion taking hold and, since a more pronounced lower temperature $GdFeO_3$ tilting favors the G-type AFM state, the onset of a distorted G-type AFM consolidates at 77 K, the temperature for the first order structural phase transition.

Further, as it was somehow anticipated above, we note that from the lattice point of view, these competitive interactions, conflicts between various ordering mechanisms resulting from faulty orbital arrangements imply instabilities that in our case will be also likely related with the Jahn-Teller distortion in the <u>ab</u> plane below 210 K. This competition causes frustration and intergrowth, yielding an incommensurate (INC) ion arrangement, i.e., long range order but no translational symmetry by not having one atomic position exactly repeated from cell to cell in at least one crystallographic direction.[16] In $YVO_3$, and although superlattice reflections have not been detected in the temperature interval from 210 to 77 K, this situation is found in orbital mismatching of the *ab* plane. This is supported by the new infrared active hard phonons resembling the overall behavior already found in incommensurate prototype materials such as $K_2SeO_4$ (ref. 17). By now it is accepted that in oxides nanoscale phase separation appearing as intrinsic lattice inhomogenieties is a current feature revealing a phenomenology similar to manganese oxides that may be even extended to cuprates.[18] In YVO3, being an insulator in the whole temperature range, provides an unusual and comprehensible window to look at this phenomena. Then, having faulty orbital ordering in the *ab* plane immediately below ~210 K, the above perspective yields an average crystal symmetry close or equal to that of the higher temperature phase and, on cooling toward 77 K, we would expect the appearance of domains that carry inside the essence of the low temperature ordered (COMM) phase. This cluster nucleation and growth would trigger the broad first order transition "locking-in" the lattice periodicity at ~77 K and it would explain the origin short range orbital fluctuations that, being weak, are not altering the overall long range ordering in a length scale of 1000 Å.[10] This proposition will also help to solve the riddle found in first principle calculations needing to endorse the persistence of the low temperature phase distortion in the intermediate phase requiring a more than 50 percent mixture of it.[7] We also add that around 77 K microdislocations mechanisms would cause the found hysteretic behavior.[10]

The magnetic order onset at ~116 K might be identified with a change in lattice regime from orthorhombic-like to a more monoclinic-like and magnetic reversal would be related to orbital quasi-order altering the canting angle locally signaling to a more sensitive magnetic probe the onset of cluster formation in a soliton regime. As it was in the case of incommensurate dielectrics we did not expect to detect nor we did observe that change of regime by optical means.[19]

On the other hand, it is worth mentioning that specific heat measurements by Boruhovich et al[20] and, more recently, by Blake et al[10] show transitions having a characteristic inverted λ shapes with a long tail on the low temperature side extending all the way to 77 K and beyond. In addition to a magnetic role in $YVO_3$, lattice contributions yield very broad transitions that are similar to what has been found for incommensurate phases in prototype $A_2BX_4$ systems[21,22] The inverted lambda shape is consistent with discommensuration, here related to a non-negligible orbital disorder, and it is opposite to the behavior found at structural and other phase transitions. We note that it has been established that in incommensurate dielectrics the normal-INC transition the enthalphy and entropy are generally much larger that of the INC-COMM transitions. On the high temperature side the heat capacity decreases rapidly.[21]

Finally, another point to remark is that YVO$_3$ is inherently highly anharmonic even at room temperature. Figure 5 shows 1 cm$^{-1}$ resolution infrared transmission measurements with defined very weak substructures in a phonon lattice band related to the sharper structure, shown in figure 1 in reflectivity, for the same frequency range. This would indicate the lattice inhomogeneities are also present at room temperature within in a local mode picture, thus passing undetected by techniques dealing with long range ordering and Rietveld analyses. In essence, they reveal a latent lower than orthorhombic P*bmn*-D$_{2h}^{16}$ symmetry even at 300 K at some spots of the YVO$_3$ lattice. On the other hand, it is then significant noting that resonant X-ray diffraction studies suggest that the G-type orbital ordering starts at room temperature.[23]

Summarizing, we have shown that new hard phonons become infrared active below 210 K. We propose that the gradual net Brillouin zone center folding may be originated from faulty orbital stacking due to local energetically prevailing distortions between 210 and 77 K. Fluctuations around an average structure may be consequence, as in dielectrics, of discommensuration regions that cooling toward T$_{N2}$ will have the intersoliton region growing in the coherence of domain walls sliding fluctuations that, exceeding some critical value, would trigger the low temperature monoclinic ordered phase. We associated this to the observed anomalous Debye-Waller factors. We conclude that the intermediate phase of $YVO_3$ provides a rich ground on novel interactions that remain to be explored.

N.E.M. acknowledges an enlightening exchange of e-mails with T. T. M. Palstra (University of Groningen, Netherlands). J. A. A., M. J. M-L and M. T. C. acknowledge the financial assistance of the Ministerio de Ciencia y Tecnología under Project Nº MAT2001-0539. C. P. also acknowledges a Ph.D. grant (PROC. 00/00789-3) to the funding agency from the state of São Paulo(FAPESP), Brazil, and the Laboratorio Nacional de Investigación y Servicios en Espectroscopía Optica, thanks the CLAF (Centro LatinoAmericano de Física) for an emergency maintenance grant.

## Figure Captions

**Figure 1**. Lattice phonons emerging between 200 and 77 K due to the folding of the Brillouin zone. Lines with filled circles mark longitudinal optical mode frequencies. The inset shows the reflectivity of $YVO_3$ at 300, 200 and 77 K. For clarity, the spectra at 200 and 77 K have been offset by an arbitrary constant.

**Figure 2.** Phonon oscillator strengths in the intermediate phase of $YVO_3$.

**Figure 3.** Fourier Transform of EXAFS signal representing the pseudo-radial distribution function around V atoms. Inset : Detail of the pseudo radial distribution function for V nearest neighbors.

**Figure 4**. Temperature dependence of the mean relative Debye-Waller factor obtained from the fitting using three different fixed V-O distances The inset explains the model used in the fit. The stars show the distances found by neutron diffraction measurements after ref.10. The open triangles are the average distances seen by EXAFS and, finally, the filled squares are the distances used in the fit for the temperatures of the EXAFS measurements. The full line is a guide for the eye.

**Figure 5**. Transmission spectra of lattice phonon side bands in 1 cm$^{-1}$ resolution spectra of YVO$_3$ between 225 and 210 cm$^{-1}$ at 300 and 77 K.

## Fitting Parameters for the Reflectivity of YVO$_3$

| T (K) | $\varepsilon_\infty$ | $\Omega_{to}$ (cm$^{-1}$) | $\Omega_{lo}$ (cm$^{-1}$) | $\gamma_{to}$ (cm$^{-1}$) | $\gamma_{lo}$ (cm$^{-1}$) | $S_j$ (cm$^{-2}$) |
|---|---|---|---|---|---|---|
| 300 | 2.45 | 90.4 | 102.6 | 546.0 | 493.5 | 3.38 |
| | | 140.6 | 142.3 | 33.9 | 37.8 | 1.05 |
| | | 175.0 | 176.8 | 7.5 | 9.5 | 0.4 |
| | | 182.7 | 195.5 | 17.1 | 10.4 | 2.3 |
| | | 203.1 | 218.9 | 4.9 | 60.9 | 0.76 |
| | | 230.3 | 242.6 | 105.6 | 39.1 | 0.38 |
| | | 248.7 | 250.0 | 12.8 | 8.4 | **0.02** |
| | | 277.6 | 283.8 | 17.2 | 21.7 | **0.32** |
| | | 320.5 | 325.5 | 18.9 | 14.0 | 0.88 |
| | | 328.7 | 359.0 | 11.2 | 39.0 | 0.50 |
| | | 360.2 | 361.4 | 6.6 | 11.4 | 0.002 |
| | | 395.2 | 403.6 | 17.0 | 22.5 | 0.53 |
| | | 408.9 | 447.7 | 12.6 | 50.5 | 0.32 |
| | | 461.9 | 507.7 | 27.4 | 32.9 | 0.17 |
| | | 561.8 | 627.1 | 15.0 | 249.0 | 0.27 |
| | | 667.9 | 668.3 | 20.8 | 18.2 | 0.003 |
| | | 677.5 | 702.6 | 61.7 | 32.7 | 0.06 |
| 200 | 2.42 | 99.1 | 103.6 | 128.4 | 113.6 | 0.98 |
| | | 134.6 | 136.7 | 40.7 | 51.3 | 0.45 |
| | | 179.4 | 186.0 | 7.7 | 20.5 | 2.01 |
| | | 190.5 | 195.9 | 12.0 | 7.10 | 0.61 |
| | | 204.6 | 220.7 | 3.41 | 31.6 | 1.10 |
| | | 231.2 | 243.6 | 41.4 | 25.3 | 0.42 |
| | | 247.8 | 250.1 | 9.1 | 7.6 | **0.03** |
| | | 277.9 | 283.6 | 14.7 | 22.5 | **0.29** |
| | | 318.8 | 324.4 | 11.6 | 9.8 | 0.82 |
| | | 328.7 | 353.4 | 6.4 | 39.1 | 0.53 |
| | | 360.4 | 364.3 | 9.2 | 13.4 | 0.05 |
| | | 395.7 | 412.5 | 15.2 | 21.3 | 0.91 |
| | | 413.1 | 446.8 | 9.6 | 55.8 | 0.03 |
| | | 461.3 | 505.5 | 25.1 | 38.3 | 0.20 |
| | | 559.3 | 629.6 | 23.5 | 25.4 | 0.29 |
| | | 657.3 | 659.6 | 23.5 | 25.4 | 0.006 |
| | | 686.5 | 706.9 | 73.3 | 24.1 | 0.05 |
| | | 979.9 | 1014.4 | 1283.2 | 2238.1 | 0.12 |
| 100 | 2.37 | 89.0 | 91.9 | 82.9 | 55.5 | 0.89 |
| | | 156.8 | 169.9 | 232.8 | 22.4 | 3.52 |
| | | 179.7 | 181.1 | 7.42 | 549.6 | 1.03 |
| | | 181.7 | 199.8 | 29.7 | 5.2 | 0.45 |
| | | 205.4 | 219.3 | 0.8 | 17.7 | 0.27 |
| | | 231.0 | 237.4 | 18.5 | 8.8 | 0.21 |
| | | 247.8 | 254.6 | 14.5 | 6.0 | **0.18** |
| | | 276.1 | 285.4 | 13.6 | 18.2 | **0.31** |
| | | 315.3 | 323.5 | 10.3 | 15.2 | 0.44 |
| | | 330.2 | 340.1 | 4.00 | 20.1 | 0.19 |
| | | 357.2 | 365.8 | 14.6 | 10.2 | 0.23 |
| | | 388.5 | 416.3 | 21.6 | 65.3 | 0.58 |
| | | 417.4 | 427.4 | 15.1 | 17.4 | 0.01 |
| | | 453.8 | 501.9 | 40.4 | 32.5 | 0.29 |
|     |      | 562.8   | 609.8   | 8.2    | 110.3  | 0.24  |
|     |      | 646.8   | 674.3   | 53.5   | 70.7   | 0.07  |
|     |      | 695.4   | 705.4   | 44.1   | 18.8   | 0.01  |
|     |      | 1027.0  | 1038.4  | 1417.1 | 2318.1 | 0.04  |
| 77  | 2.40 | 89.0    | 91.9    | 93.6   | 61.7   | 0.91  |
|     |      | 151.8   | 169.9   | 276.5  | 23.9   | 3.69  |
|     |      | 178.8   | 181.9   | 6.05   | 667.5  | 1.13  |
|     |      | 181.7   | 199.6   | 26.4   | 5.7    | 0.32  |
|     |      | 205.5   | 217.3   | 0.24   | 16.4   | 0.27  |
|     |      | 230.6   | 237.0   | 21.6   | 8.2    | 0.24  |
|     |      | 247.4   | 255.2   | 14.5   | 6.7    | **0.23** |
|     |      | 276.5   | 286.4   | 10.2   | 11.7   | **0.34** |
|     |      | 313.6   | 320.3   | 5.2    | 7.9    | 0.37  |
|     |      | 329.9   | 341.0   | 4.4    | 15.3   | 0.32  |
|     |      | 355.4   | 365.7   | 15.5   | 8.3    | 0.28  |
|     |      | 383.2   | 416.3   | 15.8   | 88.5   | 0.53  |
|     |      | 417.4   | 425.9   | 24.7   | 19.3   | 0.008 |
|     |      | 448.9   | 505.6   | 66.9   | 30.1   | 0.26  |
|     |      | 553.1   | 559.9   | 14.3   | 25.5   | 0.07  |
|     |      | 565.5   | 589.4   | 6.2    | 141.6  | 0.05  |
|     |      | 654.5   | 669.3   | 55.1   | 47.4   | 0.07  |
|     |      | 689.1   | 709.5   | 58.4   | 29.9   | 0.04  |
|     |      | 1015.9  | 1025.72 | 1290.2 | 2036.1 | 0.03  |

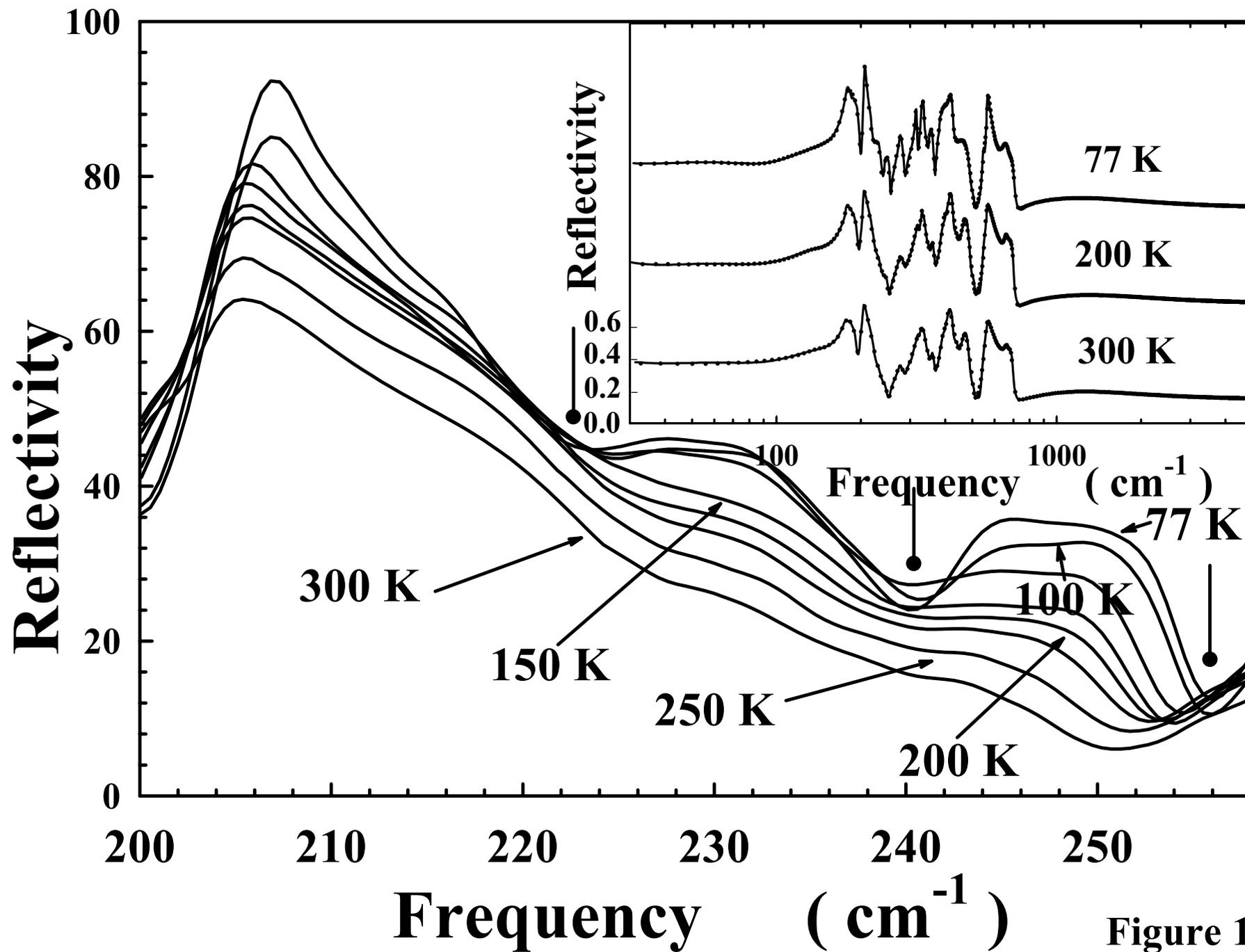

Figure 1
Massa et al

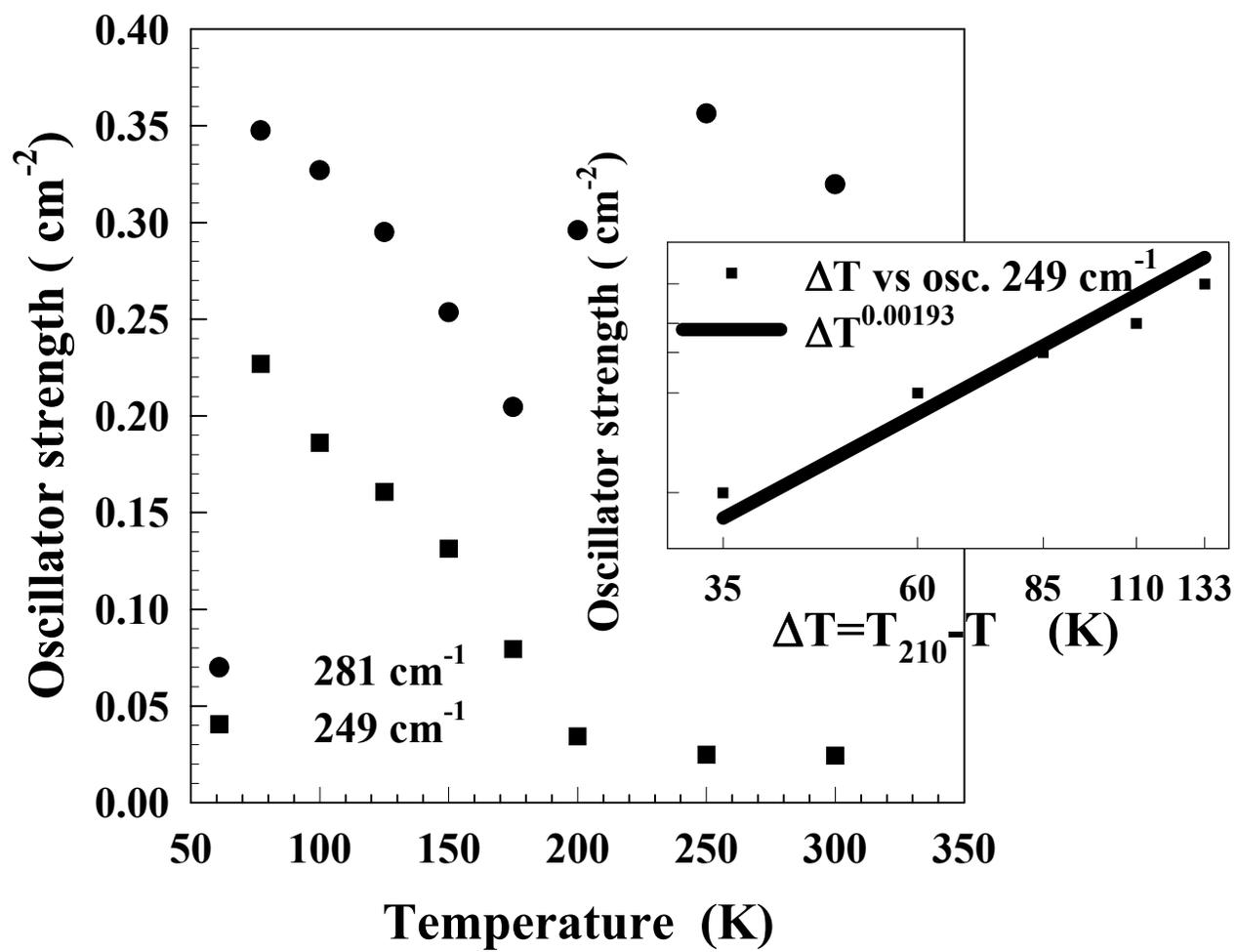

**Figure 2
Massa et al**

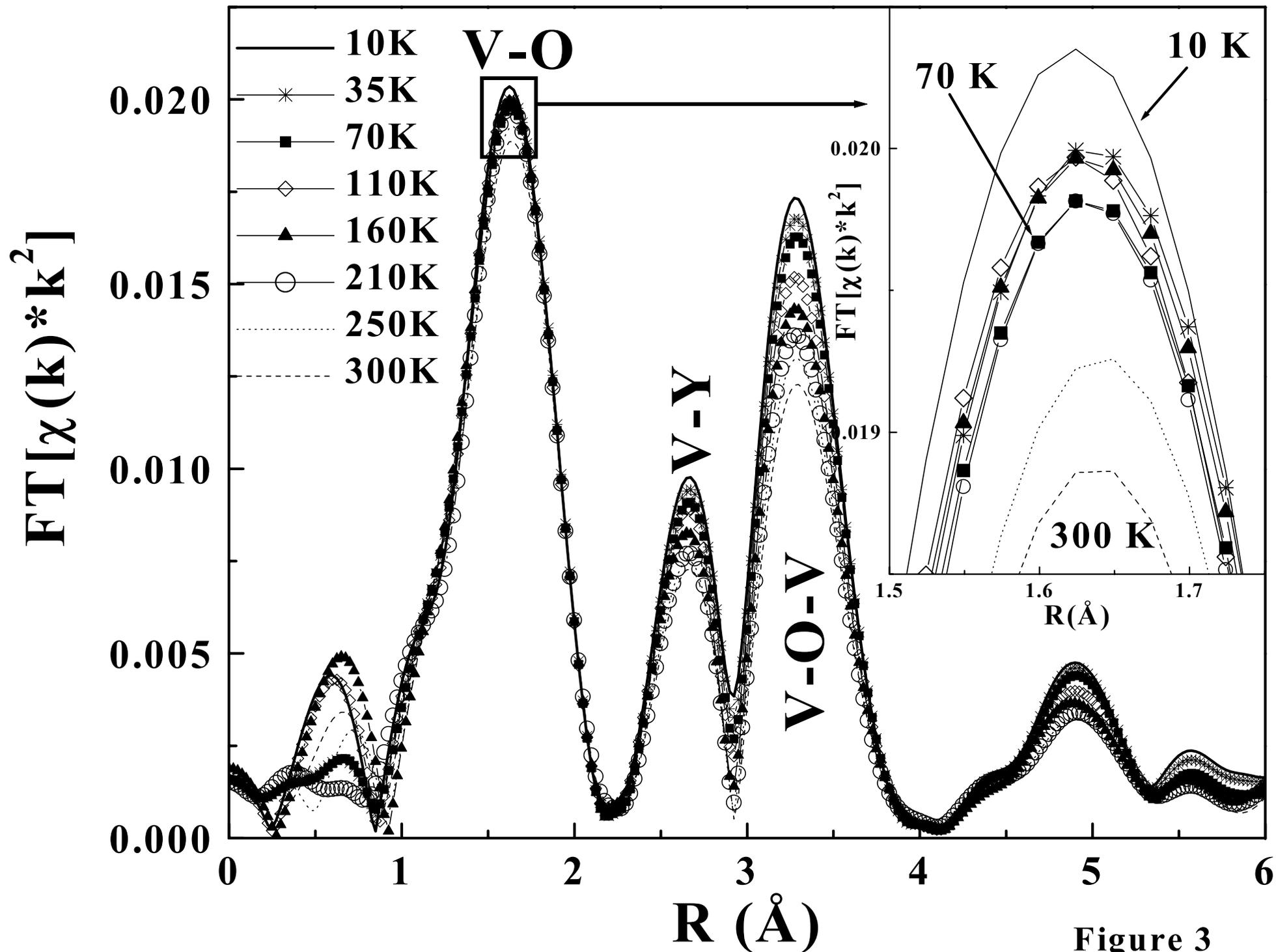

Figure 3
Massa et al

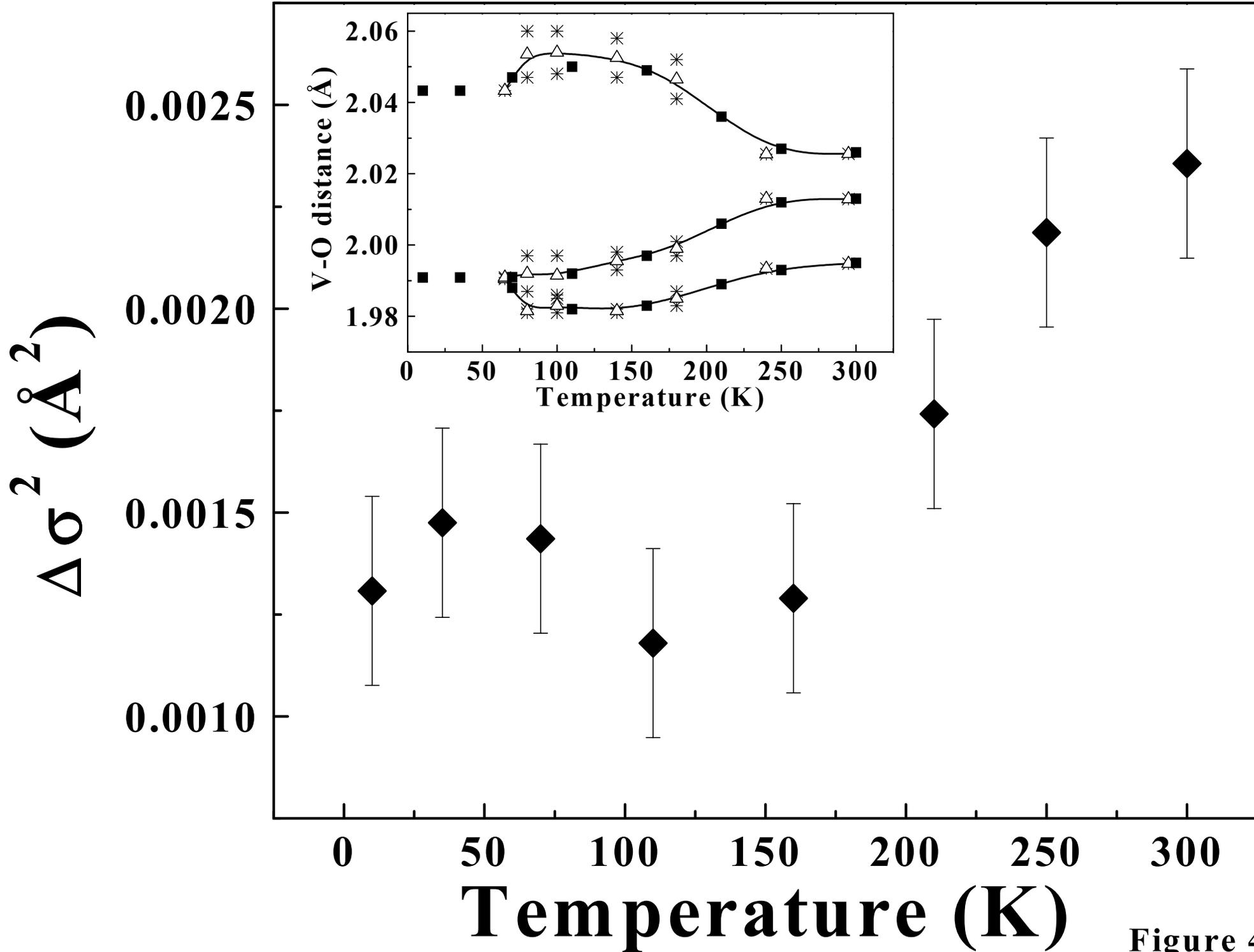

Figure 4
Massa et al.

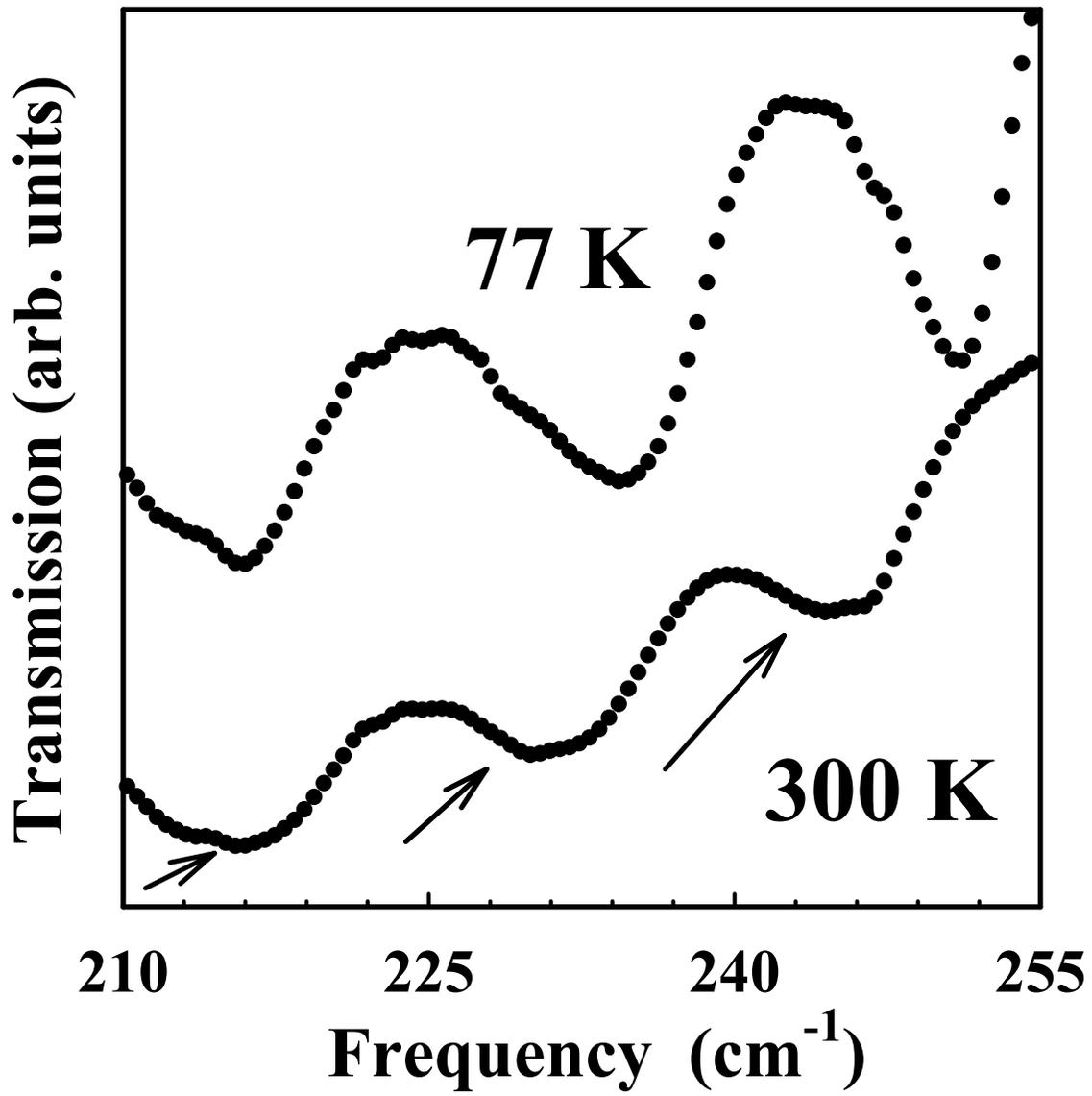

Massa et al
Figure 5